\documentclass{jpsj-suppl}
\usepackage{txfonts} %Please comment out this line unless the txfonts package is availabe in your LaTeX system.

\def\bra{\langle}
\def\ket{\rangle}
\def\rmd{{\rm d}}

\def\half#1{{#1\over 2}}

\def\crsth{{[$q^3$8\half3]}}
\def\crsh{{[$q^3$8\half1]}}
\def\cwsh{{[$q^3$1\half1]}}

\def\Lamc{\mbox{$\Lambda_c$}}
\def\Sigc{\mbox{$\Sigma_c$}}
\def\Sigcstar{\mbox{$\Sigma_c^*$}}

\def\Jpsi{\mbox{$J\!/\!\psi$}}
\def\etac{\mbox{$\eta_c$}}
\def\Dbar{\mbox{$\overline{D}$}}
\def\Dbarstar{\mbox{$\overline{D}{}^*$}}

\def\cbar{\overline{{c}}}

\def\qbar{\overline{{q}}}

\newcommand{\xbld}[1]{\mbox{\boldmath $#1$}}

\def\vecr{{\xbld{r}}}

\def\vecR{{\xbld{R}}}

\def\rmd{{\rm d}}

\def\half#1{\text{${#1\over 2}$}}

\def\Vcoul{V_\text{coul}}
\def\Vconf{V_\text{conf}}
\def\Vcmi{V_\text{cmi}}
\def\Ocmi{{{\cal O}_\text{cmi}}}

\title{The ccbar Pentaquarks by a Quark Model }

\author{Sachiko \textsc{Takeuchi}$^{1,3,4}$ and Makoto \textsc{Takizawa}$^{2,4,5}$}

\inst{$^{1}$Japan College of Social Work, Kiyose, Tokyo 204-8555, Japan \\
$^{2}$Showa Pharmaceutical University, Machida, Tokyo 194-8543, Japan\\
$^{3}$Research Center for Nuclear Physics (RCNP), Osaka University, Ibaraki, Osaka, 567-0047, Japan\\
$^{4}$Theoretical Research Division, Nishina Center, RIKEN, Wako, Saitama 351-0198, Japan\\
$^{5}$J-PARC Branch, KEK Theory Center, IPNS, KEK, Tokai, Ibaraki, 319-1106, Japan}

\email{s.takeuchi@jcsw.ac.jp}

\recdate{}

\abst{Recent LHCb experiments have shown us 
that there are two resonances in the  $\Jpsi p$ channel in the $\Lambda_b$ decay, 
whose spin and parity are most probably (3/2$^-$ 5/2$^+$). 
In this work, we investigate the $I(J^P)=\half1(\half1^-)$, $\half1(\half3^-)$, and $\half1(\half5^-)$ $uudc\cbar$ pentaquark states 
 by employing the quark cluster model.
It is found that the color-octet isospin-\half1 spin-\half3 $uud$ configuration
gives an attraction to such five-quark systems. 
This configuration together with the color-octet $c\cbar$ pair 
gives structures around the $\Sigc{}^{(*)}\Dbar{}^{(*)}$
thresholds:
one bound state, two resonances, and one large cusp are found in the $uudc\cbar$  negative parity channels.
We argue that these resonances and cusp may
correspond to, or  combine to form,
the negative parity pentaquark peak observed by LHCb.
}

\kword{hidden-charm pentaquark; color-octet baryon; exotic hadron; multiquark hadron; baryon-meson scattering}

\begin{document}
\maketitle
\section{Introduction}

In 2015, 
two candidates of the new exotic baryons, $P_c$(4380) and $P_c$(4450), 
had been reported by LHCb.
They are observed in the $\Lambda_b^0\rightarrow \Jpsi p K^-$ decay.
The $P_c$(4380) has a mass of 
4380$\pm$8$\pm$29 MeV and a width of 205$\pm$18$\pm$86 MeV
while $P_c$(4450) has a mass of 
4449.8$\pm$1.7$\pm$2.5 MeV and a width of 39$\pm$5$\pm$19 MeV.
The most favorable set of the spin parity for the lower and the higher peaks is  
$J^P = (\half3^-,\half5^+)$,
but $(\half3^+,\half5^-)$ or $(\half5^+,\half3^-)$ are also acceptable
according to their analysis \cite{Aaij:2015tga}.
Their configuration is considered to be $uudc\cbar$: a hidden-charm pentaquark
of the isospin \half1.

In this work, we discuss 
the negative-parity $udsc\cbar$ pentaquarks  \cite{Takeuchi:2016ejt}.
They are considered to couple to baryon-meson states,
and their feature is observed
in the short range properties of such baryon-meson states.
%, which is considered to be governed by the quark and gluon dynamics. 
For this purpose, we employ the quark cluster model, 
which successfully explained the short range part of the baryon-baryon interaction and
the structure of the light-flavored pentaquark $\Lambda(1405)$ \cite{Oka:2000wj,Takeuchi:2007tv}.
Recent lattice QCD results 
are found to give  similar short range potentials 
to those of the quark cluster model for the baryon-baryon interaction \cite{Sasaki:2015ab}. 

\begin{table}[t]
\renewcommand\arraystretch{1.4}
\tabcolsep=1mm
\caption{The classification of the isospin-$\half1$ negative parity $qqqc\cbar$ states.
The $uud$ spin ($s_q$), color ($c$), 
CMI of the five quark systems at the heavy quark limit 
($\bra \Ocmi \ket_{5q}^{(HQ)}$), the possible five quark spin with the multiplicity ($J$),
the lowest $S$-wave threshold (T) and the CMI contribution to the threshold energy ($\bra \Ocmi \ket_\text{T}^{(HQ)}$) are listed.}
\begin{center}
\begin{tabular}{ccccllccccccccccccccccccc}\hline
& $s_q$  & $c$  & $\bra \Ocmi \ket_{5q}^{(HQ)}$ &~~~$J$ & ~~~ T & $\bra \Ocmi \ket_\text{T}^{(HQ)}$
\\\hline
\cwsh\  &$\half1$ & {\bf 1} &$-8$ & (\half1)$^2$, \half3 &$N\etac$, $N\Jpsi$ & $-8$
\\
\crsh\ & $\half1$ & {\bf 8} &$-2$ & (\half1)$^2$, \half3 & \Lamc\Dbar$^{(*)}$&$-8$
\\
\crsth\ & $\half3$ & {\bf 8} &\phantom{$-$}2& \half1, (\half3)$^2$, \half5 & $\Sigma_c^{(*)}$\Dbar$^{(*)}$& ${8\over 3}$
\\\hline
\end{tabular}
\end{center}
\label{tbl:hq}
\end{table}%

Let us first discuss possible configurations of $uud$ quarks in the $uudc\cbar$ pentaquarks.
These three light quarks can be color-singlet or color-octet.
So, when the orbital configuration is totally symmetric,
the $uud$ configuration in the $uudc\cbar$ systems can be totally symmetric ({\bf 56}-plet)
or mixed symmetric ({\bf 70}-plet) in the flavor-spin
SU$_{f\sigma}$(6) 
space accordingly. They are classified as:
\begin{align}
{\bf 56}_{f\sigma} 
&=
  {\bf 8}_f \times {\bf 2}_\sigma
+ {\bf 10}_f\times {\bf 4}_\sigma
,~~~~
{\bf 70}_{f\sigma} 
= 
  {\bf 1}_f \times {\bf 2}_\sigma
+ {\bf 8}_f \times {\bf 2}_\sigma
+ {\bf 8}_f \times {\bf 4}_\sigma
+ {\bf 10}_f\times {\bf 2}_\sigma\ .
\end{align}
The color-singlet $uud$ systems
correspond to the usual {\bf 56}-plet baryons, whereas 
the color-octet ones correspond to the {\bf 70}-plet systems. %,
%which 
%can be decomposed into 
%the flavor-singlet spin-$\half1$ (${\bf 1}_f\times {\bf 2}_\sigma$), 
%the flavor-octet spin-$\half1$ (${\bf 8}_f\times {\bf 2}_\sigma$), 
%the flavor-octet spin-$\half3$  states (${\bf 8}_f\times {\bf 4}_\sigma$),
%and the flavor-decaplet spin-$\half1$ (${\bf 10}_f\times {\bf 2}_\sigma$).
Since the present work concerns systems of the isospin $\half1$ and the strangeness zero,
the configurations of the three light quarks correspond to  
one of the following three:
(a)  color-singlet spin-$\half1$ baryon in (${\bf 8}_f\times {\bf 2}_\sigma$), 
namely, nucleon, 
(b)  color-octet spin-$\half1$ $q^3$ in (${\bf 8}_f\times {\bf 2}_\sigma$),  and
(c)  color-octet spin-$\half3$ $q^3$ in (${\bf 8}_f\times {\bf 4}_\sigma$).
In the following,
we denote each of them by \cwsh, \crsh, and \crsth, respectively.
Since the spin of the $c\cbar$ pair is either 0 or 1, 
the total spin of the $uudc\cbar$ systems is
either $\half1$ (5-fold), $\half3$ (4-fold), or $\half5$ (1-fold). (See Table \ref{tbl:hq}.)

In Table \ref{tbl:hq}, we list the color magnetic interaction (CMI) 
evaluated by the $uud$ part of the five-quark system, ($\bra \Ocmi \ket_{5q}^{(HQ)}$), 
which corresponds to the CMI contribution to the five-quark system at the heavy quark limit.
The lowest $S$-wave thresholds (T) are also shown together with the CMI contribution to the threshold energy ($\bra \Ocmi \ket_\text{T}^{(HQ)}$).
As seen from the table, $\bra \Ocmi \ket_{5q}^{(HQ)}$ is smaller than  $\bra \Ocmi \ket_\text{T}^{(HQ)}$ for the \crsth\ configuration;
which means that CMI is attractive  in this configuration.
Since $uudc\cbar$ is color-singlet as a whole, 
the system of the color-octet $uud$ with the color-octet $c\cbar$ 
can be observed as \Lamc\Dbar${}^{(*)}$ or \Sigc${}^{(*)}$\Dbar${}^{(*)}$ baryon meson states,
where each of the hadrons is color-singlet.
The above CMI contribution is expected to be seen as an attraction in the  \Sigc${}^{(*)}$\Dbar${}^{(*)}$ baryon meson channels. 
We argue that this attraction may cause the one of the observed peaks by LHCb.

\section{Model}
The model Hamiltonian, 
$H_q$, consists of the central term, $H_c$,
and the color spin term, $\Vcmi$.
The $H_c$ consists of 
the kinetic term, $K$, the confinement term, $\Vconf$, and the 
color Coulomb term, $\Vcoul$:
\begin{align}
H_q&=H_c+\Vcmi, ~~~~
H_c= K+\Vconf+\Vcoul\ .
\end{align}
Both of the $\Vcoul$ and $\Vcmi$ terms come from the effective one-gluon exchange interaction
between the quarks.

The color flavor spin part of the $q^3$ or $q\qbar$ wave functions is
taken as a conventional way  \cite{pdg}.
The orbital wave function of the mesons, $\phi_M$, and that of the baryons, $\phi_B$,
are written by Gaussian with a size parameter $b$, $\phi(\vecr,b)$:
\begin{align}
\phi_M(\vecr_M) &= \phi(\vecr_{12},{x_{0}\over \sqrt{\mu_{12}}})
,~~~~
\phi_B(\vecr_B) 
=\phi(\vecr_{12},{x_{0}\over \sqrt{\mu_{12}}}) \phi(\vecr_{12-3},{x_{0}\over \sqrt{\mu_{12-3}}})
\ ,
\end{align}
where the reduced masses, $\mu_{12}$ and $\mu_{12-3}$, 
correspond to the Jacobi coordinates, $\vecr_{12}$ and $\vecr_{12-3}$.
We assume that the size parameter of the orbital motion can be 
approximated by $b = x_0/\sqrt{m}$ and
minimize the central part of the Hamiltonian, $H_c$, against 
$x_0$ for each of the flavor sets: 
$u\cbar$, $c\cbar$, $uud$, $udc$.
For the baryons, this means that the ratio of the size parameters is kept to a certain mass ratio;
 {\it e.g.}, $b_{uc}/b_{ud}$
in \Lamc\ or \Sigc\ is equal to $\sqrt{\mu_{ud}/\mu_{uc}}$. %$\sqrt{(m_u+m_c)/2m_c}$.

We employ the resonating group method (RGM) in order to solve the five-quark systems.
The wave function of the five quark system, $\Psi$, 
consists of the $q^3$ baryon and the $q\qbar$ meson with the relative wave function 
$\chi$
 \cite{Oka:2000wj,Takeuchi:2007tv}:
\begin{align}
\Psi &= \sum_{\nu} c^\nu
{\cal A}_q \big\{\psi_B^\nu(\vecr_B)\psi_M^\nu(\vecr_M)\chi^\nu(\vecR)\big\}
%\\
%\chi(\vecR,\vecSp_i)&=i_0({1\over b^2}\vecR\cdot\vecSp_i)\exp[-{1\over 2b^2}(R^2+S_i^2)]
\ ,
\end{align}
where ${\cal A}_q$ stands for the quark antisymmetrization which operates on the four quarks, and $\nu$ for the baryon-meson channel.
%, 
%and $i_\ell(z)$ is the modified spherical bessel function.
%%
%As for the scattering state, the wave function of the relative motion 
%is connected smoothly to the spherical Hankel functions in the long range region.
%
By integrating out the internal wave function of the hadrons, 
the RGM equation can be obtained from the equation of motion for the quarks, 
$(H_q-E) \Psi  = 0$, as
\begin{align}
\sum_{\nu'} \int (H^{\nu\nu'}-EN^{\nu\nu'}) \chi^{\nu'}=0
~,
\end{align}
where $H^{\nu\nu'}$ and $N^{\nu\nu'}$ are the hamiltonian and the normalization kernels.

In order to investigate the
nature of the resonance states as well as the bound states,
We define  a three-body operator, $P^{cs}$, 
to extract the $uud$ color $c$, spin $s_q$, orbital $(0s)^3$ component:
\begin{align}
{\cal P}^{cs_q}  &= {\cal P}^{cs_q}_{123}+{\cal P}^{cs_q}_{124}+{\cal P}^{cs_q}_{134}+{\cal P}^{cs_q}_{234}
\label{eq:P0s}
\\
{\cal P}^{cs_q}_{ijk}  &= |uud;cs_q(0s)^3\ket \bra uud;cs_q(0s)^3|
%=P^{f}_{ijk}P^\sigma_{ijk} P^c_{ijk}P^{orb}_{ijk}
%\\
%P^f_{ijk} &=\begin{cases}1 & \text{all of the $ijk$-th quarks are the light quarks.}\\ 0 & \text{else}\end{cases}
%\\
%P^\sigma_{ijk} &={1\over 6 }(\pm(\sigma_i\sigma_j+\sigma_j\sigma_k+\sigma_i\sigma_k)+3),
%\label{eq:Psijk}
%\\
%P^c_{ijk}&={1\over 6 }(\pm(\lambda_i\lambda_j+\lambda_j\lambda_k+\lambda_i\lambda_k-1)+3)
%\label{eq:Pcijk}
%\\
%P^{orb}_{ijk}(\vecxi,\vecxi')&=\phi(\vecr_{ij},b_{uu})\phi(\vecr_{ij-k},{\sqrt{3}\over 2}b_{uu})
%\phi(\vecr'_{ij},b_{uu})\phi(\vecr'_{ij-k},{\sqrt{3}\over 2}b_{uu})
%\delta^3(\vecr_{c\cbar}-\vecr'_{c\cbar})
%\delta^3(\vecR_{uud-c\cbar}-\vecR'_{uud-c\cbar})
%\label{eq:Poijk}
\\
%{\cal P}^{cs}&=\sum_{\nu\nu'}
%\int\rmd \vecr \rmd\vecr' \rmd\vecxi\rmd\vecxi'\;
%  \psi_b^{\nu\dag}(\vecr_b)\psi_m^{\nu\dag}(\vecr_m)\chi^\dag(\vecR)
% {\cal A}_q \sum_{ijk}{\cal P}^{cs}_{ijk}(\vecxi,\vecxi') {\cal A}_q 
%\big\{\psi_b^{\nu'}(\vecr'_b)\psi_m^{\nu'}(\vecr'_m)\chi(\vecR')\big\}
\bra{\cal P}^{cs_q}\ket&=\sum_{\nu\nu'}
\int\prod \rmd \vecr\;
\big\{  \psi_b^{\nu}\psi_m^{\nu}\chi^{\nu}\big\}^\dag
 %{\cal A}_q 
 \sum_{ijk}{\cal P}^{cs_q}_{ijk} {\cal A}_q 
\big\{\psi_b^{\nu'}\psi_m^{\nu'}\chi^{\nu'}\big\}~.
%\\
%&=
%\int\rmd \vecr_b \rmd\vecr_m\rmd\vecR\;
%  \psi_b^{c\dag}(\vecr_b)\psi_m^{c\dag}(\vecr_m)\phi^\dag(\vecR,b)
% {\cal O} (1-3P_{34}^{\sigma f c o}) 
%\psi_b^{c'}(\vecr_b)\psi_m^{c'}(\vecr_m)\phi(\vecR,b)
%\\
%&=
%  \psi_b^{c\dag}\psi_m^{c\dag} {\cal O}^{\sigma f c} (1-3P_{34}^{\sigma f c}) 
%\psi_b^{c'}\psi_m^{c'}
\label{eq:40}
\end{align}
%where $+$($-$) is taken for the spin \half3\ (\half1) projection operator in  Eq.\ (\ref{eq:Psijk}),
%whereas $+$($-$) is taken for the color singlet (octet) projection operator.
%$P^c_{ijk}$ is a simplified projection operator,
%knowing that the color of the $uud$ system is restricted to be singlet or octet.
%As for the size parameter of the $(0s)^3$ configuration, 
%we use the same value as that of $b_{uu}$ in $\Sigma_c$. 

\section{Results}

\begin{figure}[tbp]
\begin{center}
\includegraphics[trim = 6mm 3mm 8mm 3mm, clip, width=7.6cm]{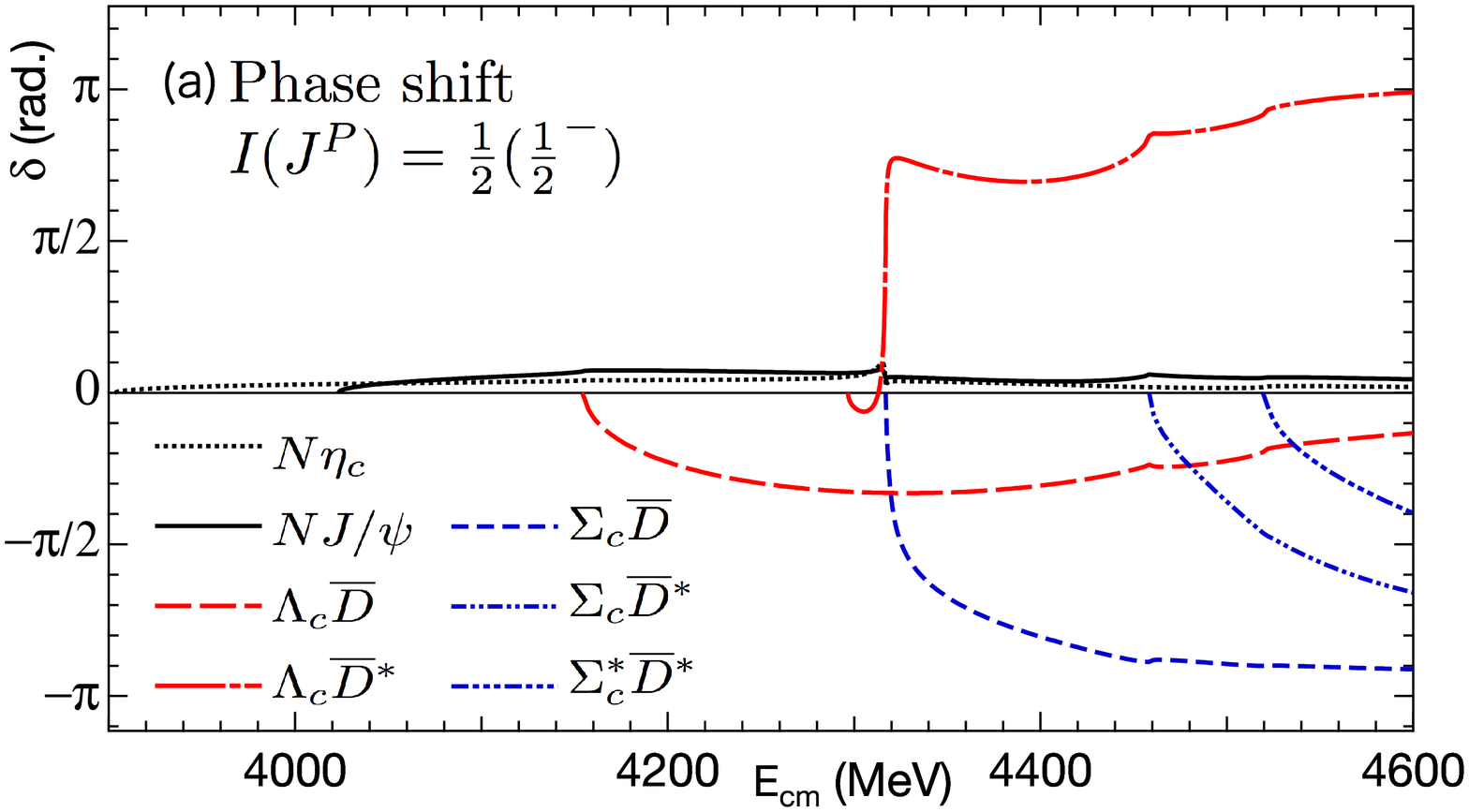}
\includegraphics[trim = 6mm 3mm 8mm 3mm, clip, width=7.6cm]{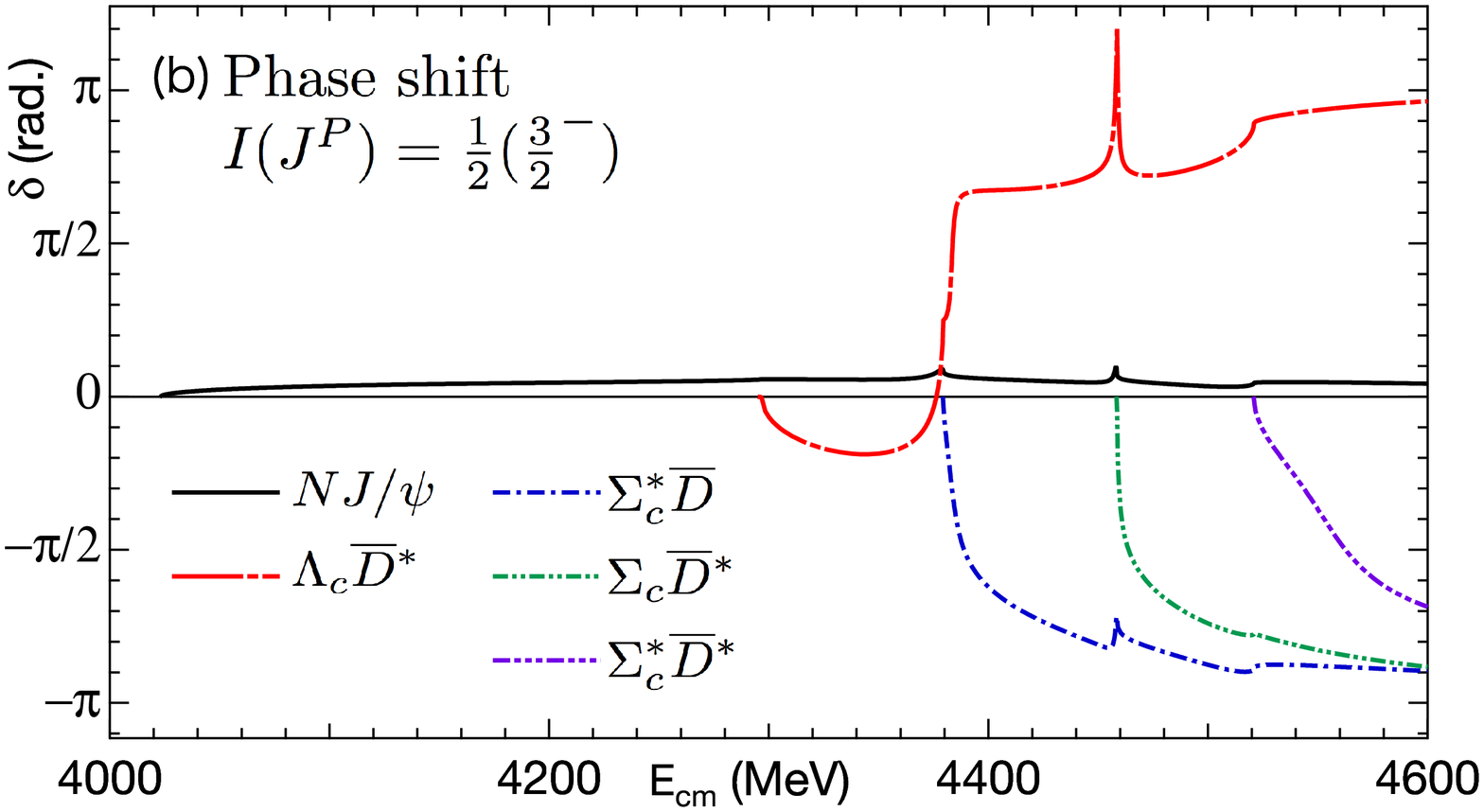}
\caption{The scattering phase shift of the $S$-wave $uudc\cbar$ $I(J^P)$=$\half1(\half1^-)$ channel (Fig.\ a) and that of the $\half1(\half3^-)$ channel (Fig.\ b).
The solid line is that of the $N\Jpsi$ channel, 
the dotted line the $N\eta_c$, the long-dashed and the long-dot-dashed lines are for 
\Lamc\Dbar\ and \Lamc\Dbarstar, and the dashed, dot-dashed, double-dot-dashed, and  triple-dot-dashed lines are 
for the  \Sigc\Dbar, \Sigcstar\Dbar, \Sigc\Dbarstar, and \Sigcstar\Dbarstar,
respectively. Figures are taken from ref.\ \cite{Takeuchi:2016ejt}.
(color online)}
\label{fig:ps}
\end{center}
\end{figure}

\begin{figure}[tbh]
\begin{center}
\includegraphics[trim = 6mm 3mm 6mm 1mm, clip, width=5.5cm]{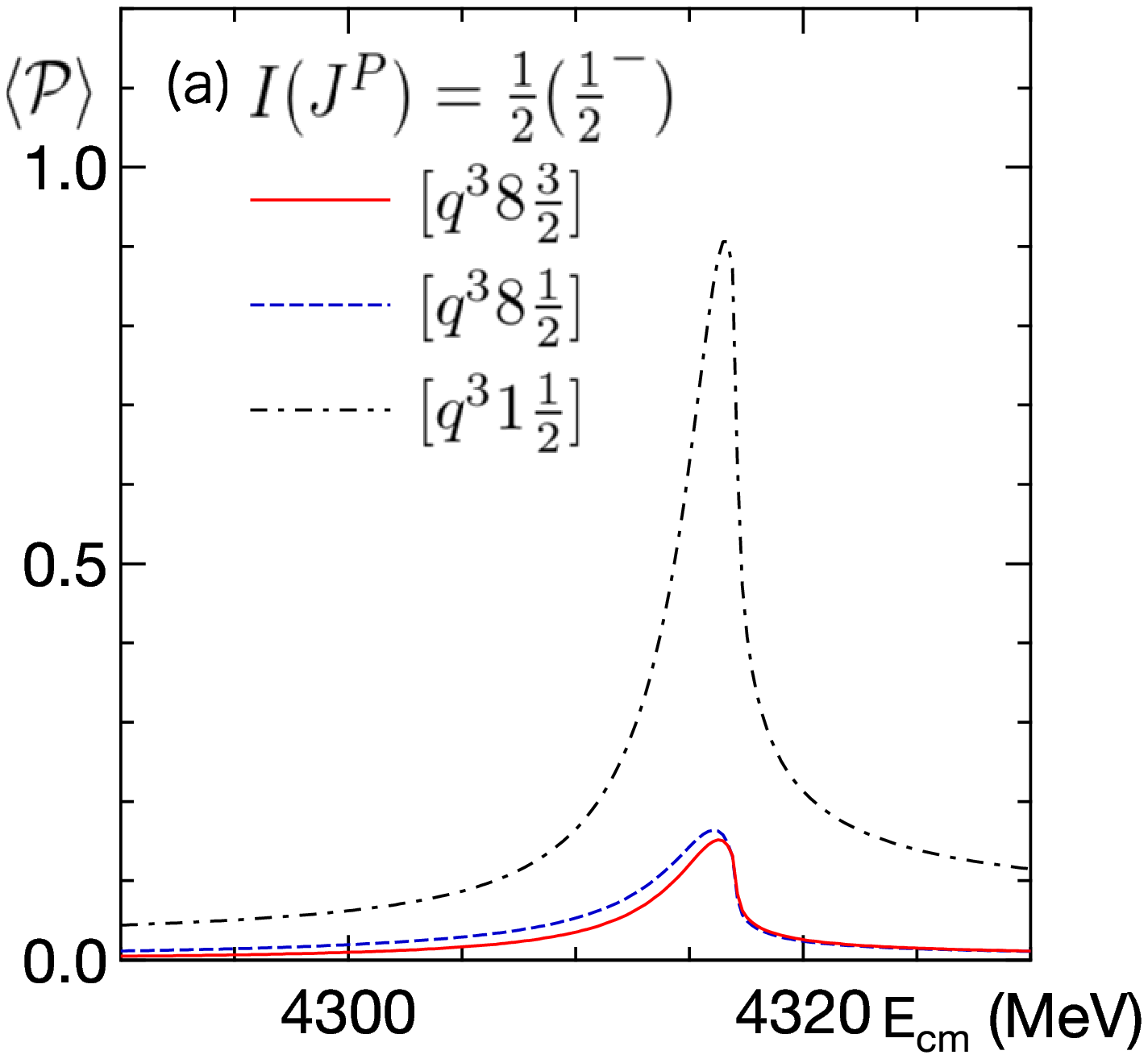}\rule{1cm}{0cm}
\includegraphics[trim = 6mm 3mm 6mm 1mm, clip, width=5.5cm]{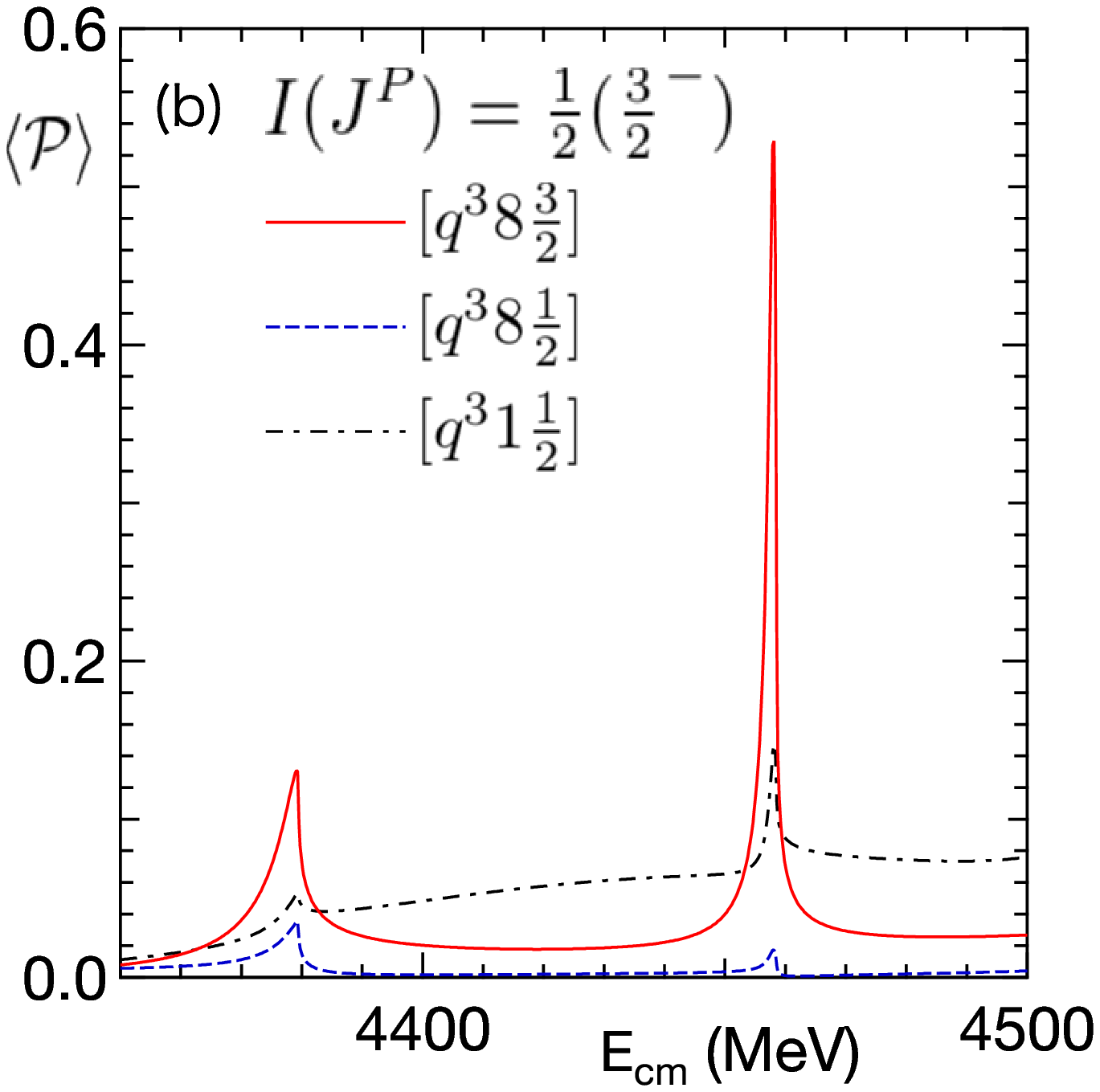}
\end{center}
\caption{The factor $\bra {\cal P}\ket $ to find the $uud(0s)^3$ configuration 
in the scattering wave function with the initial $N$\Jpsi\ channel 
in the $S$-wave $uudc\cbar$ $I(J^P)$=$\half1(\half1^-)$  channel (Fig.\ a) and that of the $\half1(\half3^-)$ channel (Fig.\ b).
The solid line stands for the factor to find in the color octet spin \half3,
the dashed line for the color-octet spin \half1, and the dot-dashed line for the color singlet spin \half1. (color online)}
\label{f1}
\end{figure}

It is found that a very shallow bound state appears
in the \Sigcstar\Dbarstar\ $J=\half5$ system,
in which the $uud$ is in the \crsth\ configuration.
As seen from the scattering phase shifts shown in Figs.\ 1 (a) and (b), 
there is one sharp resonance in the 
\Lamc\Dbarstar\ channel of the $J=\half1$ system,
while one sharp resonance and one strong cusp are found in the 
\Lamc\Dbarstar\ channel of the $J=\half3$ system.
The number of these structures, one in $J=$\half1, two in \half3, one in \half5,  
corresponds exactly to the number of the multiplicity
for the \crsth\ configuration shown in Table 1.

The $\bra{\cal P}^{cs}\ket$'s are shown in Figs.\ 2 (a) and (b). 
The size of \crsth\ configuration enhances at the resonance or the cusp energies.
This feature is more clearly found in the $J=\half3$ system.
There, the attraction 
originally forms a bound state 
each in the \Sigcstar\Dbar\ and in the \Sigc\Dbarstar,
which become a resonance and a cusp in the \Lamc\Dbarstar\ channel
by the channel coupling.
At these resonance and cusp energies, 
the short range part of the wave function enhances
and the proportion of \crsth\ also enhances.
These resonance and cusp indeed have the \crsth\  configuration at the short range part.
As for the $J=\half1$ system, the size of this configuration 
 is much smaller than that of the \half3 system.
Without the coupling to the \Lamc\Dbar\ channel, however, 
the resonance becomes a bound state of the \Sigc\Dbar\ channel, 
and the \crsth\ component is 0.7 of the whole $uud~(0s)^3$ component of that bound state.
This configuration plays an important role to make the
resonance also for the $J=\half1$ system
though the mixing of the \Lamc\Dbar\ channel reduces its size.
The above situation shows us that \crsth, the $uud$ color-octet spin \half3 configuration,
which may be called as a `color-octet $uud$ baryon,' causes these resonances.

In order to compare our results to the 
experimental spectra, it will be necessary to include
the effects of the meson-exchange in the long range baryon-meson interaction.
We would like to argue, however, these resonances and cusp may
correspond to, or  combine to form,
the negative parity pentaquark peak observed by LHCb.

\section{Summary}

The $I(J^P)=\half1(\half1^-)$, $\half1(\half3^-)$, and $\half1(\half5^-)$ $uudc\cbar$ systems 
are investigated by the quark cluster model.
It is shown that the color-octet isospin-\half1 spin-\half3 $uud$ configuration 
gains attraction from the color magnetic interaction.
The $uudc\cbar$ states with this configuration 
cause structures around the $\Sigc{}^{(*)}\Dbar{}^{(*)}$
thresholds.
We have found
one bound state in $\half1(\half5^-)$, %4520 MeV,
one resonance and a cusp in $\half1(\half3^-)$, and
one resonance in $\half1(\half1^-)$ 
in the negative parity channels,
which may be the origin of the negative parity $P_c$ peak observed 
in the $\Lambda_b$ decay.

%We would like to thank Professors M.\ Oka and
%A.\ Hosaka for useful discussions.

\end{document}